# Using Targeted Phonon Excitation to Modulate Thermal Conductivity of Boron Nitride

Dongkai Pan[1], Xiao Wan[1], Tianhao Li[1,2], Zhicheng Zong[1,2], Yangjun Qin[1,2], Nuo Yang[2*]

[1] School of Energy and Power Engineering, Huazhong University of Science and Technology, Wuhan 430074, China

[2] Department of Physics, National University of Defense Technology, Changsha 410073, China

[*]Corresponding author. Email: nuo@nudt.edu.cn

# Abstract

Recent advancements in thermal conductivity modulating strategies have shown promising enhancements to the thermal management capabilities of two-dimensional materials. In this article, both iterative Boltzmann transport equation solution and two-temperature model were employed to investigate the efficacy of targeted phonon excitation applied to hexagonal boron nitride. The results indicate significant modifications to hBN's thermal conductivity, achieving increases of up to 30.1% as well as decreases of up to 59.8%. These findings validate the reliability of the strategy, expand its scope of applicability, and establish it as a powerful tool for tailoring thermal properties across a wider range of fields.

Thermal conductivity is a critical property with significant implications across various applications. The requirements for thermal conductivity vary depending on specific circumstances. In thermal management, high thermal conductivity is essential for efficient heat dissipation to prevent device overheating and ensure reliability [1]. Conversely, in thermoelectric applications, low thermal conductivity is preferred to achieve a high figure of merit (zT) and enhance thermoelectric performance [2–4]. These contrasting demands underscore the importance of effective modulation techniques [5]. Phonon transport predominantly governs heat conduction in most semiconductors, thus phonon-based modulation strategies are extensively researched.

There are numerous physical mechanisms influencing phonon transport and offering pathways for modulation [5–7]. For instance, thermal conductivity can be enhanced by minimizing phonon-phonon scattering phase space and reducing phonon-impurity scattering rates, as observed in materials like cubic boron arsenide and low-dimensional materials [8–13]. Conversely, larger anharmonicity, increased crystal complexity, or additional scattering sources such as disorder and interfaces can reduce thermal conductivity through enhancing scattering strength [14,15]. Beyond scattering mechanisms, coherent regimes, phase transitions, and localization mechanisms are also promising avenues for modulation [16–23]. And the strength of interactions also significantly impacts thermal conductivity [24–26], while external factors such as electric fields and external pressure show potential for modulating as well [24,27,28]. However, these modulations typically lead to irreversible alterations to the original materials or structures, necessitating consideration of in-situ modulation strategies.

Inspired by the phonon weak coupling mechanism [29–32] and the realization of phonon modes excitation in experiments [33–35], a novel modulation approach utilizing targeted phonon excitation to modulate thermal conductivity has been proposed recently [36]. In low-dimensional systems, phonon-phonon interactions are much weaker compared to those observed in conventional bulk materials. This attenuation in interactions is also known as phonon weak coupling. The weak coupling mechanism limits the energy exchange between phonon modes, thereby facilitating the thermal conductivity modulation based on selective modal excitation. Its efficacy has been validated on graphene, which has a super high thermal conductivity, through ab initio calculations and molecular dynamics simulations. Moreover, results on other materials are crucial for the demonstration of its broader applicability. Here, hBN is chosen as the platform due to its widespread use as an encapsulation material and its notably lower thermal conductivity compared to graphene despite both being 2D phonon weak coupling systems.

In this work, the effect of targeted phonon excitation on thermal conductivity modulation is further explored on hexagonal boron nitride (hBN). Initially, the phonon density of states (DOS), scattering rates and spectral mapping of thermal conductivity are calculated as a guidance of subsequent excitation and modulation. Based on the interplay between various mechanisms, the modulation results are first predicted qualitatively and then validated against the calculated outcomes. Two methods are employed to calculate post-exciting modulation thermal conductivity: iterative Boltzmann transport equation (BTE) solution [37] and two-temperature model (TTM) [29]. The results from the two methods are compared to ensure their consistency. In this way, the effectiveness of the modulation strategy is validated on hBN.

The modulating outcome hinges on the selection of targeted phonon modes, with a focus on their participation in heat conduction. Several key parameters, such as scattering rates, spectral thermal conductivity, and phonon DOS, are chosen to characterize the involvement of the phonon modes within specific frequency ranges in the heat conducting process.

Scattering rate serves as a fundamental metric in estimating the resistance in transport processes, where a higher scattering rate generally indicates worse transport properties. Modes with high scattering rates are prioritized in modulation aimed at reducing thermal conductivity, as their excitation tends to adversely affect other modes.

Additionally, modulation targets all modes within certain frequency ranges rather than targeting individual modes, due to the dense distribution of modes and the difficulty in selectively exciting one mode. In this case, spectral mapping of thermal conductivity is utilized to represent the contribution of modes across different frequency ranges to heat conduction. The modes within frequency ranges exhibiting higher spectral thermal conductivity are leveraged to enhance thermal conductivity.

What's more, while phonon DOS is not directly related to the phonon transport process, it dictates the number of modes within unit frequency intervals, thus determining the number of modes for excitation. In other words, a higher DOS signifies more modes within corresponding frequency ranges, which can lead to more significant effects on thermal conductivity, regardless of whether the overall impact is positive or negative.

Building upon the established modulation principles, the computation details are presented below. A $5 \times 5 \times 1$ supercell was constructed for the calculation of second and third-order force constants. The second-order force constants were obtained via density functional perturbation theory, implemented by the Vienna Ab initio Simulation Package (VASP) [38,39], while the third-order force constants were calculated using the finite displacement method. And the cutoff radius for the calculation of 3rd-order force constants was set to be the 6th nearest neighbor.

The overall thermal conductivity is calculated via the iterative solution of linearized phonon BTE with a $\Gamma$-centered $50 \times 50 \times 1$ regular grid at a temperature of 300K (a temperature of 300 K is assumed throughout this paper unless otherwise specified). Modulation is implemented by introducing a multiplier (N) to the phonon distribution function, as expressed in equation (1). This modification is incorporated into the ShengBTE code [40] to realize the excitation of targeted phonon modes. The incorporation is achieved by inserting the multiplied distribution function into the calculations of scattering rates and heat capacity. To illustrate this approach, the formula for calculating three-phonon scattering rates for emission processes is presented in Equation (2). Distribution functions within the selected frequency range are multiplied with N. Here, a specialized version of ShengBTE developed by Ruan et al [41] is utilized as the baseline for its enhanced convergence performance on two-dimensional materials. And the consideration of anharmonicity is up to the third order.

$$f = \frac{N}{\exp(\hbar\omega/k_B T) - 1} \tag{1}$$

$$\Gamma^-_{\lambda\lambda'\lambda''} = \frac{\hbar\pi}{4} \frac{f'_0 + f''_0 + 1}{\omega_\lambda \omega_{\lambda'} \omega_{\lambda''}} |V^-_{\lambda\lambda'\lambda''}|^2 \delta(\omega_\lambda - \omega_{\lambda'} - \omega_{\lambda''}) \tag{2}$$

Given the nonequilibrium state induced by phonon mode excitation, defining temperatures for the targeted modes specifically becomes challenging. Consequently, adjustments are made to the heat capacity calculation, which conventionally relies on the derivative of the phonon distribution function with respect to temperature. Here, the heat capacity is corrected based on the post-exciting modulation modal energy to more accurately reflect the system's thermodynamic properties as shown in equation (3).

$$T' = \frac{\hbar\omega}{k_B}\ln^{-1}(\frac{\hbar\omega}{E_\omega}+1) \tag{3}$$

As to the two-temperature model (TTM), it is well-suited for describing non-equilibrium conditions between different phonon groups in weak coupling systems, encompassing interactions such as electron-phonon and phonon-phonon couplings [29,30,42,43]. In the TTM framework, the excited phonon group and other intrinsic phonon group are treated separately, and the non-equilibrium is assumed to exist mainly between the two groups due to their weak-coupling. The TTM model parameterizes the weak coupling between the two phonon groups with the coupling coefficient $G_{ex,in}$, defined as equation (4). Here $c_{ex}$ and $c_{in}$ denote the specific heat capacities of the excited and intrinsic phonons, respectively, and $\tau_{ex,in}$ represents the relaxation time for the coupling between the two phonon groups. The equivalent thermal conductivity $\kappa$ of the system within the TTM framework is expressed as equation (5), where $\gamma = \sqrt{G_{ex,in}\left(\frac{1}{\kappa_{ex}}+\frac{1}{\kappa_{in}}\right)}$, $x$ represents the characteristic size of the system, and $\kappa_{ex}$ and $\kappa_{in}$ represent the thermal conductivity of excited and intrinsic phonons, respectively.

$$G_{ex,in} = \frac{c_{ex}c_{in}}{\tau_{ex,in}(c_{ex}+c_{in})} \tag{4}$$

$$\kappa = \frac{\kappa_{ex}+\kappa_{in}}{1-\left[\frac{c_{ex}}{c_{ex}+c_{in}}-\frac{c_{in}\kappa_{ex}}{(c_{ex}+c_{in})\kappa_{in}}\right]e^{-\gamma x}} \tag{5}$$

By utilizing two distinct approaches, the robustness and accuracy of the results could be increased. As will be revealed in the results section, the results from the two parallel approaches show good consistency and demonstrate the great modulation effect of the strategy.

As shown in Fig. 1, the scattering rates and density of states of phonon modes in h-BN are calculated as reference for further modulation. The results depicted in Fig. 1 reveal two features. Firstly, the scattering rates of ZA modes exhibit a gradual increase with frequency until reaching the top of ZA branch, which is also the first peak (Van Hove singularity) in phonon DOS. This indicates the potential of the ZA modes near the top of the ZA branch for the modulation aimed at reducing thermal conductivity. Secondly, the maximum scattering rate occurs near the second peak of phonon DOS, indicating that the modes around this frequency can also be utilized to lower the thermal conductivity. Despite the highest peak of phonon DOS falling within the range of optical branches, these modes contribute little to thermal conductivity due to their low group velocity.

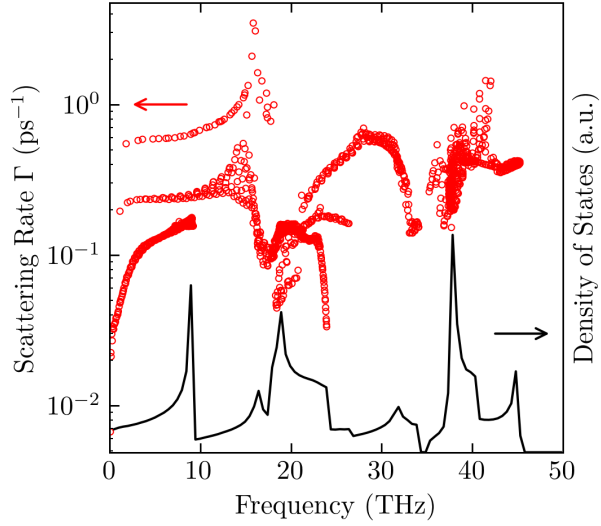

Fig. 1. The scattering rates $\Gamma$ and phonon DOS of h-BN. The red circles represent scattering rates of phonon modes, while the black line represents phonon DOS.

The spectral mapping of thermal conductivity in h-BN is calculated by differentiating the cumulative thermal conductivity, as shown in Fig. 2. The contribution from low frequency modes is surpassing compared to others, indicating the great potential of these modes for enhancing thermal conductivity. This suggests that these modes are the primary contributor to thermal conductivity. Therefore, by exciting the modes within this range, the thermal conduction with BN is supposed to be increased. At the same time, considering the scattering rates and phonon DOS in Fig. 1, it is noted that the low phonon DOS near $\Gamma$ point imposes limitations on the enhancement. Consequently, the peak for the final enhancing modulation in heat conduction should be located slightly away from the $\Gamma$ point to allow for higher phonon DOS and enough spectral mapping of thermal conductivity. Following the peak, the spectral mapping of thermal conductivity declines sharply while scattering rates increase rapidly. This causes the enhancing modulating effect to diminish fast and soon transitions to a reducing effect, which continues to intensify until the formation of a trough corresponding to the phonon DOS peak at the top of ZA branch (around 9 THz).

A similar situation is also observed at higher frequency around 18 THz, where the scattering is much stronger and spectral mapping of thermal conductivity is limited due to lower group velocity. So, the phonon DOS at this region also brings a modulation trough. As for higher optical modes, the extremely low spectral mapping of thermal conductivity of optical modes again validates their negligible contribution to thermal conductivity, so the excitation targeting them should cause no significant effect.

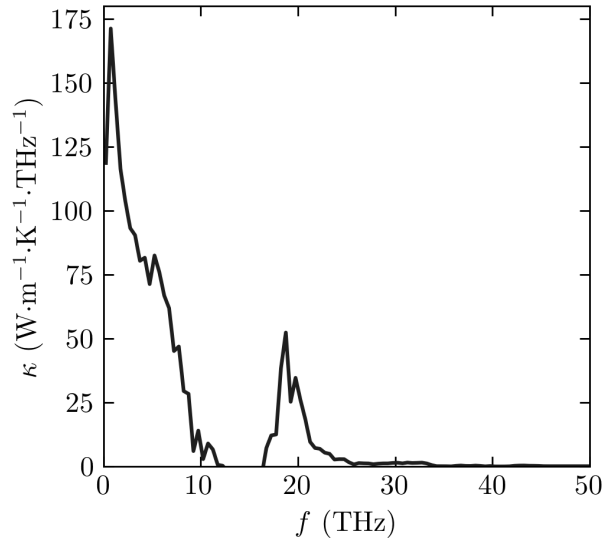

Fig. 2. The spectral mapping of thermal conductivity in h-BN. The value represents the contribution of the modes of certain frequencies to thermal conductivity. The overall thermal conductivity is 885 W·m$^{-1}$·K$^{-1}$.

Based on the above theoretical analysis, it is predicted that there will be one peak (enhancement) and two troughs (reduction) in the spectra. The peak is located close to the Γ point. Meanwhile, the first trough corresponds to the top of the ZA branch, and the second trough lies around the second peak of phonon DOS. These predictions will be validated using the subsequent calculation results.

The modulation results under different multipliers are demonstrated in Fig. 3a. The relative thermal conductivity ($\kappa/\kappa_0$) is the ratio of post-exciting thermal conductivities ($\kappa$) to the intrinsic one ($\kappa_0$). By comparing the modulation results with multiplier 5 (blue squares) and multiplier 25 (orange circles), it is evident that only amplitudes of the modulation differ, while the qualitative performance remains consistent. This confirms the robustness of this modulation strategy. Besides, in Fig. 3b, by comparing the calculation results of iterative BTE (orange circles) and two temperature model (blue triangles), it is found that the two methods are highly consistent. This further verifies the reliability of the results.

It's clearly demonstrated that there are one peak and two troughs in the results, consistent to theoretical prediction above. The peak is located at 1.45 THz (exciting the modes within 1.4−1.5 THz), while the troughs are located at 9.05 THz and 18.55 THz respectively. The peak arises from the combined effect of dominating spectral mapping of thermal conductivity and increasing phonon DOS. And the troughs result from the sharp DOS peaks and high scattering rates.

To demonstrate the feasibility of the strategy, an estimate of the energy needed for excitation is made. When the multiplier is set to 25 and phonon modes within 1.4-1.5 THz range are targeted, the required energy density is approximately $1.6 \text{ J m}^{-2}$, which is comparable to the typical energy densities achieved in experimental setups [44].

Moreover, the same pattern, containing one peak and two troughs, exists similarly in the results on graphene [36]. This similarity between results on graphene and on h-BN indicates that the modulation strategy is suitable to both two materials and hopefully, to most of the 2D materials with phonon weak coupling mechanism.

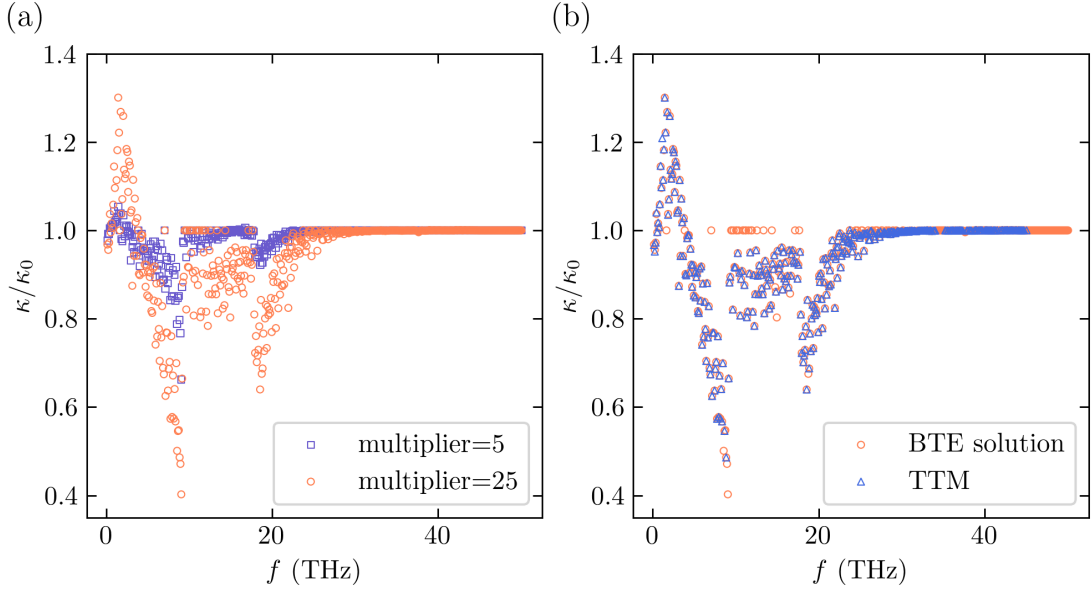

Fig. 3. Relative overall thermal conductivity of hBN ($\kappa/\kappa_0$) versus the center frequency of the excited targeted phonons. (a) Results for different multipliers. (b) Results calculated from TTM as well as iterative BTE solution when the multiplier is set to 25. Each point in the diagram represents one modulation result. For instance, the first point is located at 0.05 THz, meaning modes in 0−0.1 THz are excited with a 5 or 25 times larger energy, depending on the multiplier.

To conclude, thermal conductivity modulation based on targeted phonon excitation is further validated by iterative BTE solution and two-temperature model on h-BN. The selection of targeted phonon modes is based on a comprehensive consideration of phonon DOS, scattering rate as well as spectral mapping of thermal conductivity. And the consistent results from both parallel methods demonstrate the validity of the modulation on hBN. The results demonstrate that thermal conductivity modulation on h-BN can achieve up to a 30.1% increase and a 59.8% decrease.

These findings underscore the reliability of the strategy. The similar pattern observed in different materials suggests that this modulation strategy has broad applicability across different materials, which will provide a fantastic in-situ way to manipulate thermal conductivity. Moreover, such a quantum-level modulation can provide excellent platform for the observation of the response of macroscopic properties like thermal conductivity to microscopic perturbations. By enabling precise control over thermal conductivity at the quantum level, the strategy could significantly improve thermal management in electronic devices, enhance the efficiency of thermoelectric materials, and contribute to the development of next-generation heat control technologies.

## Author Declarations

### Conflict of Interest

The authors have no conflicts to disclose.

### Author Contributions

**Dongkai Pan:** Conceptualization (equal), Data curation (equal), Formal analysis (equal), Investigation (lead), Methodology (equal), Writing – original draft (lead), Writing – review & editing (equal)
**Xiao Wan:** Conceptualization (equal), Data curation (supporting), Formal Analysis (equal)
**Tianhao Li:** Data curation (equal), Methodology (equal), Investigation (supporting), Writing – original draft (supporting), Writing – review & editing (equal)
**Zhicheng Zong:** Methodology (supporting), Investigation (supporting)
**Yangjun Qin:** Validation (supporting), Investigation (supporting)
**Nuo Yang:** Conceptualization (equal), Formal analysis (equal), Funding Acquisition (lead), Project administration (lead), Writing – review & editing (equal)

## Data Availability

The data that support the findings of this study are available from the corresponding author upon reasonable request.